# Visualization of Object Oriented Modeling from the Perspective of Set theory

Poornima. U. S., Suma. V.

*Abstract*— Language is a medium for communication of our thoughts. Natural language is too wide to conceive and formulate the thoughts and ideas in a precise way. As science and technology grows, the necessity of languages arouses through which the thoughts are expressed in a better manner. Set Theory is such a mathematical language for expressing the thought of interest in a realistic way. It is well suited for presenting object oriented solution model, since this implementation methodology analyzes and modulates the requirements in a realistic way. Since the design flaws are one of the factors for software failure, industries are focusing on minimizing the design defects through better solution modeling techniques and quality assessment practices. The Object Oriented (OO) solution space can be visualized using the language of Set theory with which the design architecture of modules can be well defined. It provides a strong base to quantify the relationships within and between the modules, which is a mode for measuring the complexity of solution design of any software projects. This paper provides a visualization of OO modeling from the perspective of Set theory. Thereby, it paves the path for the designers to effectively design the application which is one of the challenges of a project development. Further, this mode of visualization enables one to effectively measure and controls the design complexity leading towards reducing the design flaws and enhanced software quality.

## I. INTRODUCTION

Project success resides on the realistic approach of its development. Object oriented development methodology is a popular development methodology which observes and models the problem statement of complex systems in a realistic way. The design architecture of such methodologies holds the representative of real world entities such as classes and objects. It is observed that the design flaw increases along with the system complexity [1]. Thus, the design of a class and their relationship is a decisive factor for accessing design quality of any complex system. There is no thumb rule for solution design. A well cohesive class and less dependency of classes is one of the design quality measuring criteria for the architect. Quality of complex solution design is thus based on the experience and cognitive power of an architect. Computer Science, which is an applied field, is always appropriate to view, interpret and model the solution design using the concepts of core domains such as mathematics, for better assessment of the design output.

Set theory is a mathematical language that supports the modelling of concepts in a practical approach. It is well-suited for modelling the OO concepts which provides a new horizon for Object Oriented Modelling. Hence, an application which is developed using OO concepts can also be viewed from the set theory perspective. It is a known fact that an application design need not always be monolithic and hence will comprise of various modules in different ways, upon integration need to yield a simple and flexible design for future use.

An application design is always viewed from two perspectives, namely, static and dynamic. In static design hierarchy, a system is comprised of set of subsystems which in turn has a collection of modules. Each module contains a set of packages, classes which are interrelated as required by the solution. The dynamic design of the application provides information about the interactions that occurs between the classes. However, the set theory also encompasses of properties and relations which can be co-related to static and dynamic structure of object technology. This offers a new way of visualising the solution design of OO methodology using set theory language and thereby accessing the dependency among the modules as a measure of design complexity.

## II. LITERATURE SURVEY

The correlation of Computer Science with other fields is essential to make it more expressive and useful. To improve the quality of software product, the quality contribution from design phase, through well defined modules and their interrelations are ongoing research in the field of Computer Science.

Authors of [2] recommend the application of mathematical based evidences to real problems on an industrial scale. They express that mathematics is the only way through which complexities can be detected and resolved.

Author of [3] states that mathematical modeling supports two objectives namely i) to prove the truth of the research and ii) to support the research work through a sequence of mathematical relations.

Author in [4] emphasizes that mathematical proof enables one to describe the process by assigning numbers or symbols to attributes such as processes, product and resources of real world entities. According to him, such an assignment preserves intuitive and empirical observations about the attributes. He thus expresses that mathematical model specifies relation between theory and empirical observations.

Authors of [5] focus on the quality of solution domain design for achieving overall quality the software. The paper briefs on the metrics for different phases of software development which measures the quality of process.

Author in [6] states that validity of metrics can be established through mathematical models.

Hence, authors of [7] have introduced new quality metrics to enhance the quality of software.

Authors of [8] have further applied mathematical model to predict, control and the measure the desired level of quality.

Author of [9] highlights on considering Set theory as mathematical modeling language. It provides a wide perspective of Set theory as a simple mathematical language.

Authors in [10] propose design quality metrics for software design quality. They suggest that the design quality has a direct influence on the quality of final product.

### III. VISUALIZATION OF OBJECT ORIENTED MODELING USING SET THEORY

Set is a collection of 'things', 'entities', in a real world which falls in a vicinity. The elements of a set are grouped based on a simple formal language which involves variables, constants, operators and possible functions on set elements. If S represents a set of real numbers,

$$S = \{r_1, r_2, r_3, \ldots r_n\} \qquad ST_{Eq.1.}$$

then S can be formally defined as

$$S = \{r \mid r > 0\}$$

where the rule r>0 claims the elements of the set S. Similarly, using set theory as a mathematical modeling language, the elements of OO solution space such as modules, packages and classes are visualized as individual sets which frames the static design of an application. The module M in a solution space can be expressed as a set

$$M = \{p_1, p_2, p_3, p_4\} \qquad OO_{Eq.1.}$$

where $p_1, p_2, p_3, p_4$ are packages within a module M. However a package P in turn is a set of classes which is mathematically represented as

$$P = \{c_1, c_2, c_3\} \qquad OO_{Eq.1.1.}$$

where $c_1, c_2, c_3$ are classes in P and at the dynamic design level, a collection of objects $o_1, o_2, o_3$ constitutes an object set for a given class C.

$$C = \{o_1, o_2, o_3\} \qquad OO_{Eq.1.2.}$$

The elements of module set M or class set C is finite. The size of such sets is time dependent; set M size varies as a result of module enhancement and set C size depends on dynamic objects creation.

#### A. Object Oriented Design as an Axiomatic Set Theory

The formation of a set is not all about set theory. The set of elements generally exhibits some properties called as axioms which characterizes the set. These axioms logically prove the set properties. For instance, existence of a set with no elements is equally true as existence of a singleton set S with elements in the equation $ST_{Eq.1.}$. A set with no elements, termed as empty set ∅, upholds the empty set axiom.

$$(\varnothing)\,(\exists x)\,(\forall y)\ y \neq x \qquad ST_{Eq.2.}$$

Such emptiness or NULL property has its own significance in object oriented programming. A class or object can hold nothing when they are defined. A null class defines a class with no data and functions and provides a generic platform for the designers to create a suitable design pattern. If C is a class with no elements, it is defined as

$$C = \{\,\} \qquad OO_{EQ.2.}$$

Power set axiom of Set theory provides a view to create a set of all possible subsets from a given set S.

$$P(S) = \{x : x \in S\} \rightarrow x \subseteq S \qquad ST_{Eq.3.}$$

Given $S = \{a, b\}$, $P(S) = \{\varnothing, \{a\}, \{b\}, \{a, b\}\}$ $\qquad ST_{EQ.3.1.}$

This axiom provides an opportunity for an OO designer to form a set of all possible subclasses present in a package. It is useful at the later stage of design to assess the logical connectivity between the classes which is one of the measuring factors of the design complexity.

From $OO_{Eq.1.1.}$ the power set of a package P represented as

$$P(P) = \{\varnothing, \{c_1\}, \{c_2\} \ldots \ldots \ldots \ldots \{c_1, c_2, c_3\}\} \quad OO_{Eq.3.}$$

These subsets serve as input for measuring the class interdependency within or among package classes.

The axiom of pairing is a basis for visualizing a class as a set at static design level. The singleton axiom proposes a set as collection of elements. However, pairing axiom exhibits the property that a set can be a collection of exactly two unordered pairs of sets.

$$S = \{A, B\} \qquad ST_{Eq.4.}$$

where $A \neq B$. However, a class in a package at static design level can also be visualized as a set since it contains collection of data and functions which are logically interconnected.

$$C = \{D, F\} \qquad OO_{Eq.4.}$$

where $D = \{d_1, d_2, d_3, d_4\}$ and $F = \{f_1, f_2, f_3\}$ are logically related sets.

#### B. Applying Set Relations to Class Relationships

The sets in the universe are not individually application orient. A relation between the elements of one set (domain) is established with the elements of other set (range) based on the relevance. Basically, it is a relationship between set of inputs and set of possible outputs. In other way, the relation is a subset of ordered pairs from all possible ordered pairs which is derived from Cartesian products of two sets. If S and T are two sets where $S = \{a, b, c\}$ and $T = \{d, e, f\}$, then the Cartesian product between S, T is

$$S \times T = \{(a, d), (a, e), (a, f), (b, d), (b, e), (b, f), (c, d), (c, e), (c, f)\} \qquad ST_{Eq.5.}$$

The elements in a set can be in any order, however, in the Cartesian product, the order of the element in a pair cannot be altered. For instance, for a set $S = \{a, b, c\}$ is same as $S = \{c, a, b\}$ as far as a set S identity is defined, but, in S X T, the pair $(a, d) \neq (d, a)$.

Using this unique property of a set with S X T, a relation R between S and T is defined by taking ordered pairs in S X T as input.

$$R = \{(x, y) \mid x \in S, y \in T, x R y\} \qquad ST_{Eq.6.}$$

Thus, each ordered pair (x, y) in R satisfies the relation x R y between the elements of S and T. The subset element (x, y) in desired relation R which is derived out of S X T has some properties. These properties define whether the pairs in a relation are transitive, reflexive or symmetric in nature.

In conjunction with Object Oriented scenario, the set relation is much related to the classes or packages relationships present in a module. These elements do not serve the higher level services to the users when they stand individual. But, when related with suitable relationships such as Inheritance, Association and Aggregation, the module becomes rich in services in terms of scalability, reusability and maintainability [11][12]. To address the relationship between package elements, the possible pairs of elements are obtained by referring the equations $OO_{Eq.1.1.}$ and $ST_{Eq.5.}$, for a package P is

P X P= { (c1,c1), (c1,c2), (c1,c3), (c2,c1), (c2, c2), (c2, c3), (c3, c1), (c3, c2), (c3, c3)}    $OO_{Eq.5.}$

Thus, the Cartesian product, the ordered pairs and set properties of set elements, together provide a way to the architect to define and relate the module elements such as classes for a better solution design.

*C. Properties of a Relation and its relatedness with class relations*

When applied the properties of a set relation to different relationships between classes, it holds true.

*1) Transitive property and Inheritance Relationship*

For the ordered pairs (x , y) and (y, z) of S X T, the relation R between the pairs x R y, y R z holds true, then the relation R is transitive in nature.

$(\forall x, y) \{(x, y \in R) \text{ and } (y, z \in R)\} \rightarrow \{(x, z) \in R\}$    $ST_{Eq.7.}$

This property proved for the relationship inheritance which represents 'sharing' of data and services present in a class. Inheritance is a concept of generalisation /extension where super class is generic in nature and subclass poses the inherent property of super class along with its own property. The relationship inheritance for a package P with possible pairs in $OO_{Eq.5.}$ is visualized as

$R_I = \{(c_i, c_j) | c_i \in P, c_j \in P, c_i \text{ Inherits } c_j\}$    $OO_{Eq.6.}$

The relation $R_I$ is only transitive in nature. In the hierarchy of three classes, the super class c1 inherits to subclass c2 and c2 to c3.

(c1, c2) $\in$ R and (c2, c3) $\in$ R $\rightarrow$ (c1, c3) $\in$ R    $OO_{Eq.7.}$

The properties such as reflexive and symmetric do not hold true since c1 cannot be derived from c1 and a super class is not derived from a subclass respectively.

*2) Symmetric property and Association Relationship*

For the ordered pairs (x , y) of S X T, the relation R holds true for x R y = y R x, then the relation R is symmetric in nature.

$(\forall x, y) \{(x, y \in R)\} \rightarrow \{(y, x) \in R\}$    $ST_{Eq.8.}$

Relate to OOD, all the classes in a module are not statically interrelated like inheritance. Objects of one class could associate with another class which is of different nature. Such association provides a different logical interrelationship between two different classes to make the module rich in services. This relation, namely association relationship, is mathematically visualized as

$R_{AS} = \{(c_i, c_j) | c_i \in P, c_j \in P, c_i \text{ Associates } c_j\}$    $OO_{Eq.8.}$

The relation $R_{AS}$ is symmetric in nature. When a class c1 is logically associated with another class c2, the reflection of relationship from c2 to c1 is symmetric. However, it does not support transitive and reflexive properties.

*3) Reflexive property*

For an ordered pair (x, x), the relation R is reflexive, if x R x is true. This property is defined as

$(\forall x) (x \in S) \rightarrow (x, x) \in R$    $ST_{Eq.9}$

This property creates a mirror image of the element in a given set. From the equation $OO_{Eq.5.}$, the ordered pairs (c1,c1), (c2, c2) and (c3, c3) are reflexive pairs.

*4) Aggregation Relationship*

Another relationship in OOD, aggregation, supports 'whole-part' relationship between the classes. It reduces the design complexity of a class which big in size. The commonly used data and services are defined as separate class and whenever needed, its object can be made as a data part of another class. This relationship can be mathematically defined as

$R_{AG} = \{(c_i, c_j) | c_i \in P, c_j \in P, c_i \text{ Aggregates } c_j\}$    $OO_{Eq.9.}$

The relation $R_{AG}$ is not satisfying the any of the properties.

Table.1. Correlation between set relations and Class relationships

| Properties/Relation | Reflexive | Symmetric | Transitive |
|---|---|---|---|
| Inheritance | No | No | Yes |
| Association | No | Yes | No |
| Aggregation | No | No | No |

*D. Set Functions as basis for Coupling among Classes.*

A function is a mapping between elements of two different sets. Each element in a set S must have an image in set T so that there is a mapping between S and T.

S= {a, b, c}    T= {x, y, z}

where S is a set of inputs and T is a set of outputs.
When $f$ is function between S and T, it is represented as

$f: S \rightarrow T$    $ST_{Eq.10.}$

The input for the function $f$ is an element from the set S and the output forms the set T.

$f(a)=x$, where $a \in S$ and outputs $x \in T$.

The function $f$ on each element in S thus yields only one output element in T, but there can be same outputs from the different elements of S for the function $f$.

The Venn Diagrammatic representation of the functional relationship S and T projects the sensitivity relationship between the sets.

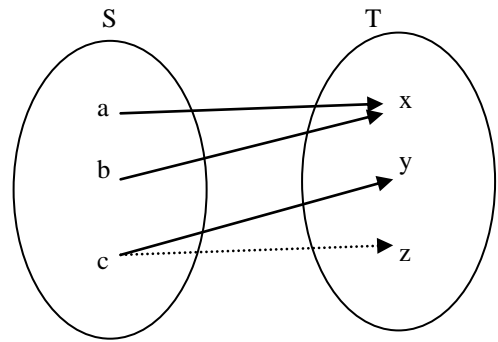

Fig.1. Venn diagram representing the functional relationship between sets.

In the Venn diagram, the element 'c' of set S has more than one output for the function $f$. So, the relationship between S and T with respect to 'c' is invalid and remains as a simple relation but not a function.

However, a function with different inputs can produce same output element.

$f(a)=f(b) =\{ x\}$  or $f^{-1}(x)= \{a, b\}$        $ST_{Eq·10.1}$

Functions in set theory can be visualized in two ways in OO scenario. Firstly, in Object Oriented Modeling, the solution space contains data, functions which is collectively called a class at static level design and objects at dynamic state of the application.

C = { {D}, {F} }

M = { {C}, {O}}        $OO_{Eq·10}$.

There is a mapping of input to output with proper function. The function transforms the input data set to output set. The functions perform the desired functionality on the input set whenever called.

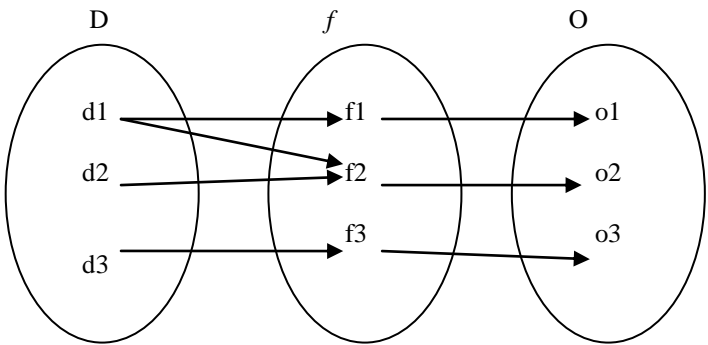

Fig.2. Venn diagram representing the functional relationship in OO class design

In other way, the classes of different nature are interrelated through sharing the services, namely 'using' relationship at dynamic phase of the system. The service of one class is requested to execute the object of another class. Such sharing of services creates coupling between the classes at run time. This relationship is essential but must be under control to reduce the complexity of the system.

## IV. COUPLING, COHESION AND METRICS

The system complexity grows when modules are added or removed. Design of such solution space when co-related with Set Theory notion, the designer would have a better understanding of the complexity of the solution design. The complexity is merely due to the amount of relationship between the elements within a class (Cohesion) and also between the classes/packages (Coupling). The Set Theory notion provides a clear view of relating the elements as well as opens up a new way of framing the metrics which are the quantitative measures of overall design quality.

## V. CONCLUSION

Increasing the success rate of defect free software is a main motto of all software development companies. One of the key phases, for minimizing the defect, is the design phase which is expected to be understandable and flexible for future changes. The challenges incurred during the design of complex software needs to be addressed in order to develop high quality software.

Mathematical modeling is proven to be one of the most effective modes of developing software products. This paper therefore presents a mathematical visualization of object oriented modeling using set theory concepts. This approach of mathematical perception when applied on design leads to development of precise software product. In addition, it is possible for designers to control measure and manage the growing design complexities due to the advancement in technology.

This paper limits to provide the prophecy of object oriented modeling using the properties and relations of set theory. From the aforementioned insight, it is now possible to develop metrics and models which can be applied at all phases of software development in order to resolve inherent software complexities.

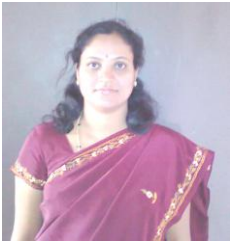 Poornima U S has done B.E. M.Tech in Computer Science and Engineering and pursuing Ph.D at Visweswaraya Technological University, India, in Computer Science and Engineering. She is currently working as an Assistant Professor in the department of Computer Science and Engineering at Raja Reddy Institute of Technology, India. She is an associate member of Research and Industry Incubation Centre, Dayananda Sagar Institutions, India and has presented several papers at International Conferences.
Email: uspaims@gmail.com

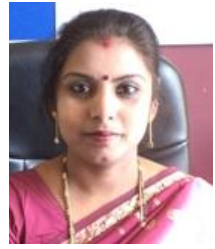 Dr. Suma. V has obtained her B.E. in Information Science and Technology, M.S. in Software Systems and Ph.D. in Computer Science and Engineering. She is currently, Dean and Head of Advanced Software Engineering Research Groups in Research and Industry Incubation Centre, Dayananda Sagar Institutions, India. She has published several International publications, which include IEEE, ASQ, Crosstalk, IET Software, ACM, IJPQM, etc. Her research results are published in NASA, UNI trier, Microsoft, CERN, IEEE, ACM portals, Springer and so on. She is awarded with several best research paper awards at various international conferences. She is an invited author for International Book Chapter. Her area of interest includes Software Engineering, Cloud Computing, Data Mining, and Information Systems.
Email: sumavdsce@gmail.com